# Spin-transfer torque switching at ultra low current densities

Johannes Christian Leutenantsmeyer*[1,2] Vladyslav Zbarsky*[1,2] Marvin von der Ehe,[2] Steffen Wittrock,[1] Patrick Peretzki,[3,4] Henning Schuhmann,[3,4] Andy Thomas,[5] Karsten Rott,[5] Günter Reiss,[5] Tae Hee Kim,[6] Michael Seibt,[3,4] and Markus Münzenberg[2]

[1] *I. Physikalisches Institut, Georg-August-Universität Göttingen, Friedrich-Hund-Platz 1, D-37077 Göttingen, Germany*

[2] *Institut für Physik, Ernst-Moritz-Arndt-Universität Greifswald, Felix-Hausdorff-Straße 6, D-17489 Greifswald, Germany*

[3] *IV. Physikalisches Institut, Georg-August-Universität Göttingen, Friedrich-Hund-Platz 1, D-37077 Göttingen, Germany*

[4] *CRC 1073, Georg-August-Universität Göttingen, Friedrich-Hund-Platz 1, D-37077 Göttingen, Germany*

[5] *Thin Films and Physics of Nanostructures, Universität Bielefeld, Universitätsstraße 25, D-33615 Bielefeld, Germany*

[6] *Department of Physics, Ewha Womans University, Seoul 120-750, South Korea*

(Dated: 27 April 2015)

The influence of the tantalum buffer layer on the magnetic anisotropy of perpendicular Co-Fe-B/MgO based magnetic tunnel junctions is studied using magneto-optical Kerr-spectroscopy. Samples without a tantalum buffer are found to exhibit no perpendicular magnetization. The transport of boron into the tantalum buffer is considered to play an important role on the switching currents of those devices. With the optimized layer stack of a perpendicular tunnel junction, a minimal critical switching current density of only 9.3 kA/cm$^2$ is observed and the thermally activated switching probability distribution is discussed.

*These authors contributed equally to this work

# I. INTRODUCTION

Spincaloric and spintronic effects in Co-Fe-B/MgO based magnetic tunnel junctions (MTJs) gained interest in recent research. Higher storage density, lower power consumption and faster access times are expected advantages. Current induced magnetization dynamics provide the opportunity to further enhance the storage density and performance of these devices [1,2,3,4].

This paper is focused on optimizing the MTJ layer stack for perpendicular magnetization anisotropy and lowering the critical current ($I_c$) for spin-transfer torque (STT) switching. A low switching current is desired for high performance storage devices [1]. Switching current densities ($J_c$) of in-plane MTJs are in the range of $10^6$ A/cm$^2$. MTJs with a perpendicular magnetic anisotropy (PMA) promise the reduction of the critical switching current while maintaining a high thermal stability Δ. Ikeda et al. reported an average $J_c$ of 3.9 MA/cm$^2$ with a switching current $I_c$ of 49 μA for PMA MTJs [3].

# II. SAMPLE FABRICATION

The MTJ stack, grown on thermally oxidized silicon substrates, consists of Ta (15 nm) / Co$_{20}$Fe$_{60}$B$_{20}$ (1.0 nm) / MgO (0.84 nm) / Co$_{20}$Fe$_{60}$B$_{20}$ (1.2 nm) / Ta (5.0 nm) / Ru (3.0 nm). Energy dispersive X-ray spectroscopy (EDX) measurements revealed a Co to Fe ratio of 32:68. Tantalum and Co-Fe-B are magnetron sputtered at an operating pressure of 1.3·10$^{-3}$ mbar Ar (base pressure approx. 5·10$^{-10}$ mbar). The MgO barrier and ruthenium capping layer are e-beam evaporated in an interconnected chamber at pressures below 5·10$^{-10}$ mbar. 5 nm tantalum are evaporated in this chamber prior to sample preparation to enhance the growth and remove moisture from the residual gas atmosphere (RGA). Afterwards, the RGA contains only N$_2$. The partial pressure of other molecules, such as H$_2$O, are below the detection limit of the quadrupole mass spectrometer (Pfeiffer QMA 200) of 10$^{-12}$ mbar. To crystallize the Co-Fe-B electrodes, the prepared samples are annealed in a separate chamber at a base pressure below 5·10$^{-6}$ mbar for 60 minutes at 300°C.

The samples are patterned using standard UV and electron-beam lithography as well as argon ion milling techniques. The junctions are of circular shape with nominal diameters between 100 nm and 250 nm. The top contact consists of 6 nm Cr and 63 nm Au. Top (Au) and bottom electrode

(Ta) are insulated with 50 nm $SiO_2$. A device is shown by Transmission Electron Microscopy (TEM) in Fig. 1. TEM and electron energy loss spectroscopy (EELS) analysis have been performed on an aberration corrected FEI Titan 80-300 ETEM G2 at 300 kV. Due to the slope of the element (see Fig. 1), resulting from the argon ion milling process under an angle to avoid redeposition, the radius of the top contact increases from 75 nm to 180 nm at the bottom where the junction is located.

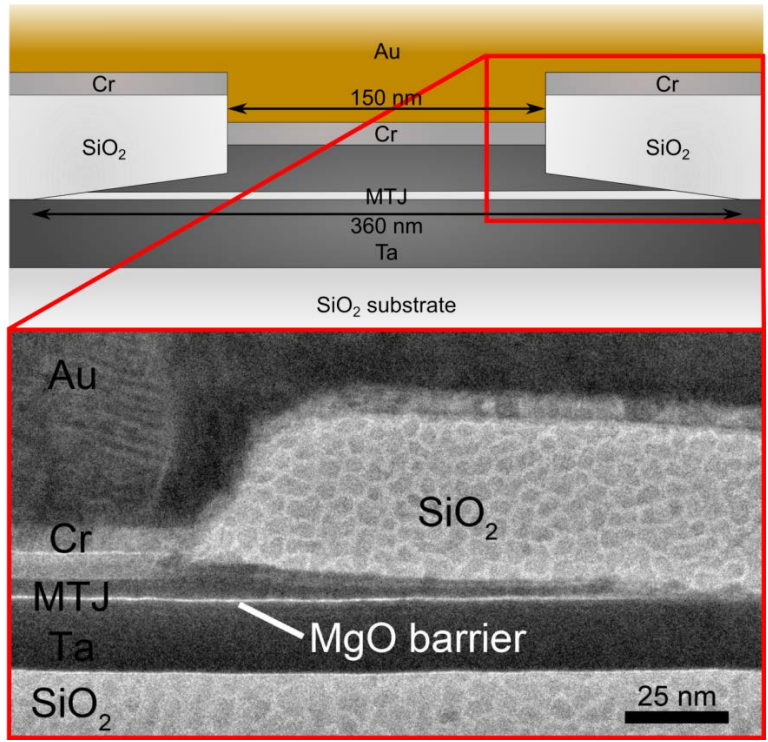

**Figure 1:** TEM data showing the cross section of a patterned PMA MTJ with a diameter of 360 nm ($Co_{20}Fe_{60}B_{20}$ (1.0 nm) / MgO (0.84 nm) / $Co_{20}Fe_{60}B_{20}$ (1.2 nm)).

## III. THE INFLUENCE OF THE BORON TRANSPORT ON THE PERPENDICULAR MAGNETIC ANISOTROPY

The interaction of the Co-Fe-B layer with the adjacent layers is crucial for the perpendicular magnetic anisotropy (PMA) whose origin is predicted to be the hybridization of the Fe 3d and O 2p orbitals at the Co-Fe-B/MgO interface layers [5]. That makes a deeper study of the influence of the Co-Fe-B/MgO interface on the magnetic behavior essential for obtaining a high PMA. In this context, Co-Fe-B/MgO based samples with a Co-Fe-B wedge were fabricated and their magnetic behavior was systematically studied.

The annealing step is crucial for the crystallization of the Co-Fe-B layer, because the Co-Fe-B crystallizes beginning at the MgO interface, in this case in bcc structure, which is essential for the coherent tunneling process [ 6]. During the annealing step, boron can diffuse from the Co-Fe-B layer into the adjacent layers and influences the crystallization of the Co-Fe. Especially the boron transport into the Co-Fe-B/MgO interface can decrease the perpendicular magnetic anisotropy of the Co-Fe layer by changing of the crystallization behavior of Co-Fe [ 7].

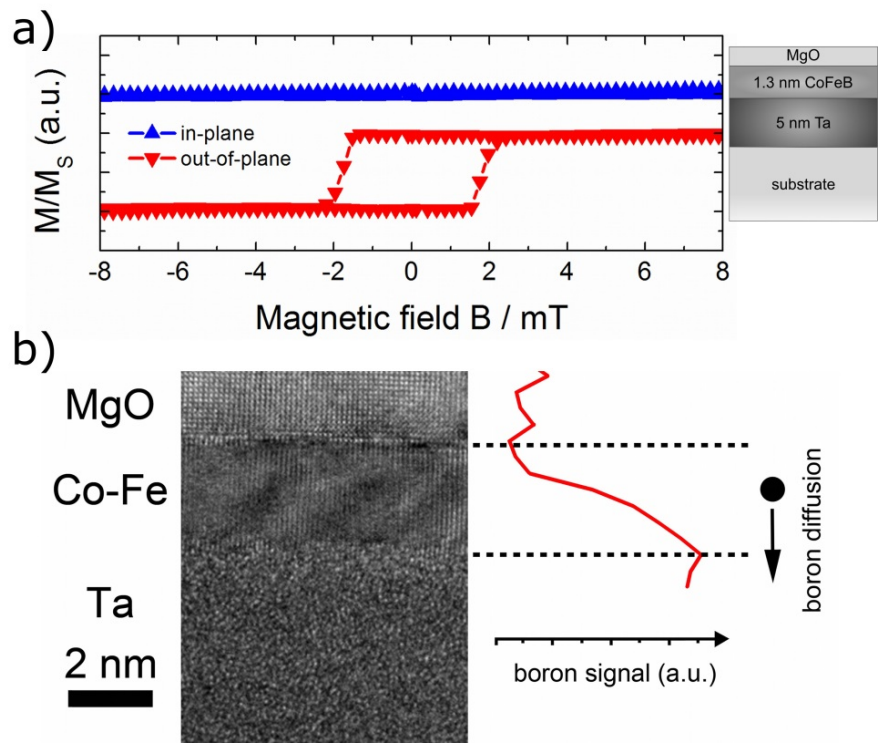

**Figure 2:** a) Hysteresis for a 1.3 nm CoFeB layer, measured for magnetic field applied in the in-plane (top) and out-of-plane (bottom) direction ($Co_{20}Fe_{60}B_{20}$ (1.33 nm)/ Ta (5.0 nm)) with MgO capping layer. b) HRTEM (left) and boron concentration after annealing (right) measured by electron energy loss spectroscopy (EELS). The concentration grows towards the CoFeB/Ta interface with a maximum in the Ta layer.

In purpose to study the magnetic behavior of the samples, the Kerr rotation (Fig. 2a)) (low field) and SQUID [ 8] (high field) measurements were carried out in the in-plane and out-of-plane directions for single CoFeB films to characterize the magnetic anisotropy defining the thermal stability in the Stoner model. The in-plane and the out-of-plane hysteresis are shown for a 1.3 nm CoFeB layer in the low field range ($Co_{20}Fe_{60}B_{20}$ (1.33 nm) / Ta (5.0 nm) with MgO cap layer). The out-of-plane hysteresis shows a squared easy-axis-loop with a coercive field of 1.8±0.02 mT, while the in-plane hysteresis is not saturated within that field range shown here (saturation field $B_S \approx 500$ mT). Using the hysteresis curves, we estimate an anisotropy constant from the calculation of the energy of the switching process. From a thickness dependent study in the large field range (not shown), we extract for the bulk anisotropy $K_b = 4.0(7) \cdot 10^5$ J/m$^3$ and for the interface anisotropy $K_i = 5.4(9)$ mJ/m$^2$, comparable to the results of Ikeda et al. [ 3]. Out-of-plane anisotropy is found below a critical layer thickness of 1.33 nm CoFeB.

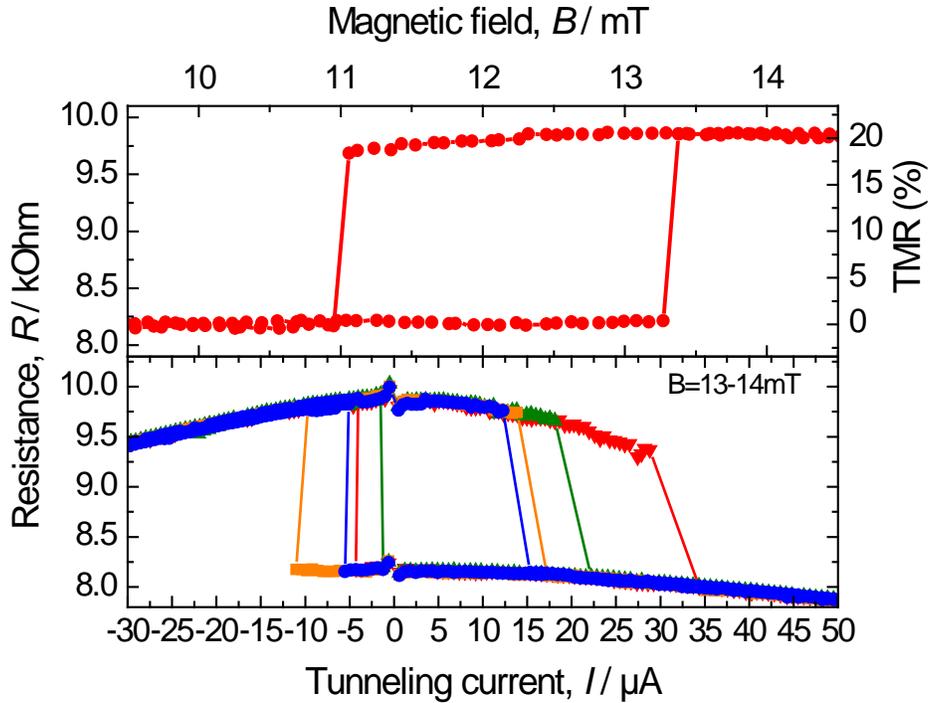

**Figure 3:** Electrical characterization of a junction. The minor loop (top) yields a TMR ratio of 20% with a perpendicularly applied magnetic field. The bottom graph shows four subsequent R(I) measurements at bias fields of 13-14 mT to illustrate the stochastic nature of the switching events. For the measurement shown as blue circles, the magnetization states are switched from P to AP alignment at an applied current of 13.5 ± 1.5 µA (corresponds to 13.3 ± 1.5 kA/cm$^2$). AP to P switching is found at -5.3 ± 0.3 µA (5.2 ± 0.3 kA/cm$^2$).

The tantalum layer influences the magnetic anisotropy behavior of the Co-Fe-B. Thus, EELS measurements were made and analyzed. The right part of Fig. 2b) shows the area of the background-corrected EEL spectrum in the range between 180 to 210 eV as a measure of the boron content. As we can see, the largest signal of boron is located in the tantalum layer and the lowest in the MgO layer. This measurement shows that boron moves into the tantalum layer during the annealing step. As a consequence, the tantalum buffer acts as a sink for boron and prevents the boron transport into the MgO layer [9,10,11] so that a higher perpendicular anisotropy can be observed due to crystallization of the Co-Fe at the MgO interface. It is known that the tantalum as buffer layer material is an essential parameter for the MTJs, because of the influence on the boron transport and thus as an important parameter for obtaining a high PMA, which our experiments support.

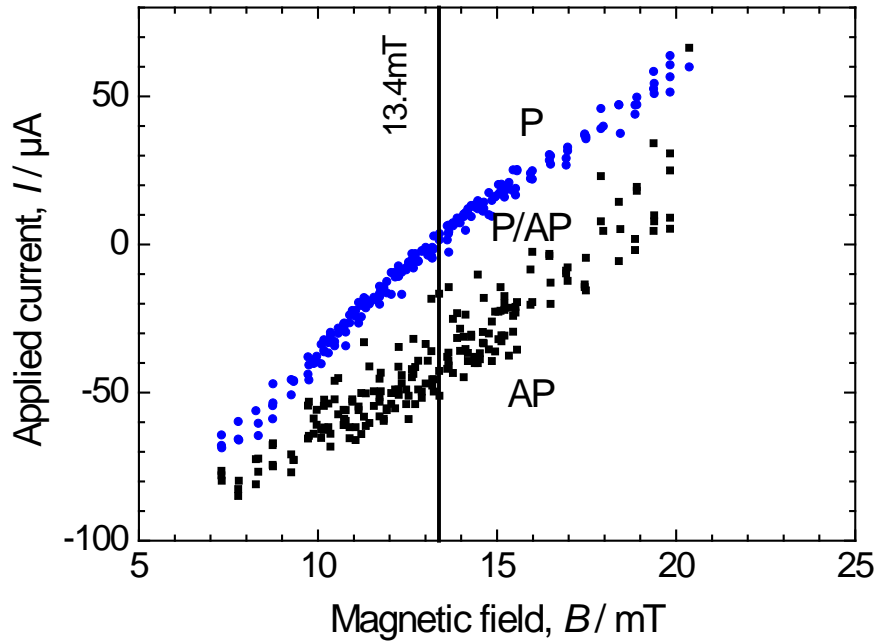

**Figure 4:** Switching phase diagram of a PMA MTJ with a diameter of 360 nm ($Co_{20}Fe_{60}B_{20}$ (1.0 nm) / MgO (0.84 nm) / $Co_{20}Fe_{60}B_{20}$ (1.2 nm)). The scattering shows the stochastic nature of the switching events and their probability distribution. The position of the spin-transfer torque (STT) data in the field range presented in Fig. 3 is marked by the vertical black line.

## IV. ULTRA LOW CRITICAL SWITCHING CURRENTS IN PERPENDICULAR TUNNEL JUNCTIONS

Here, we discuss a junction in which the smallest $J_c$ was achieved. First, the magnetoresistive behavior of the junctions is characterized. The magnetic minor loop shown in Fig. 3 is recorded in two-terminal geometry by sweeping the external magnetic field, which is applied perpendicularly to the sample plane. We find TMR ratios of up to 64% in typical PMA MTJs with a barrier thickness of 4 monolayers (ML). For this junction a TMR ratio of 20% is obtained.

Current induced switching is studied by recording the current-voltage-characteristics and is shown in the bottom graph of Fig. 3. The switching between parallel (P) and antiparallel (AP) alignment occurs at $I_c$ of -5.3 ± 0.3 µA and 13.5 ± 0.3 µA at a bias field of 13.4 mT, as indicated in the bottom graph of Fig. 4. The average of both critical switching currents (9.4 µA) equals a $J_C$ of only 9 ± 2 kA/cm². Furthermore, the influence of the bias field on the switching currents is studied. The bias-field dependent measurements yield the switching phase diagram depicted in Fig. 4. For bias fields between 7.3 mT and 20.4 mT, STT switching is observed. The STT data from Fig. 3 is marked by the vertical black line. The phase diagram shows that around a bias field of 13.4 mT, chosen in Fig. 3, the critical currents for AP→P switching are minimal. The scattering of the data points shows the stochastic nature of the switching events and their probability distribution.

In the following, we deduce average switching current densities by analyzing the switching probabilities in repeatedly recorded I-V-characteristics. Fig. 5a) shows the relative occurrences of AP→P switching events as a function of applied current through the MTJ. As a continuous line a Gaussian distribution matched to the distribution is shown. From that the AP→P switching probability distribution, plotted in Fig. 5b), is obtained by numerically integrating the Gaussian distribution of Fig. 5a). As critical current $I_c$ we define the current at which the switching probability is ≥ 95%. For AP→P switching, we obtain $I_c^{AP \to P} = -9.5$ µA. The relative frequency and probability distribution for P→AP switching is depicted in Fig. 5c) and Fig. 5d), respectively. From these data, we get $I_c^{P \to AP} = 28.2$ µA. The corresponding critical current densities are $J_c^{AP \to P} = -9.3$ kA/cm² and $J_c^{P \to AP} = 27.7$ kA/cm². To compare our results with the $J_c$ from Ref. 11 we calculate the average of the currents for P→AP and AP→P switching with $J_c = \frac{J_c^{P \to AP} + |J_c^{AP \to P}|}{2}$. Given the values stated above, we obtain $J_c = 18.5$ kA/cm². Assuming the smaller diameter of 150 nm in case of an inhomogeneous current distribution, $J_c$ would be in the order of 100 kA/cm². Values in this range were only reported by voltage induced switching, where a voltage pulse is used to reduce the energy barrier during the switching process [4].

The probability distributions of Fig. 5 allow the estimation of the thermal stability of our MTJs. According to Refs. 12 and 13, the probability distribution $P_{P(AP)}$ is given by

$$P_{P(AP)} = 1 - \exp\left(-\frac{\tau_p}{\tau_0} \cdot \exp\left[-\Delta_{P(AP)}\left(1 - \frac{I}{I_{c0}^{P(AP)}}\right)\right]\right)$$

in which $\tau_0$ is the switching duration and $\tau_p$ is the current pulse duration. This formula still agrees with Néel-Brown relaxation in presence of spin-transfer torque [14]. The energy barrier and consequently the thermal stability can be deduced from the Stoner-Wohlfarth model for a single domain or rotation in unison, respectively. For buckling or inhomogeneous magnetization reversal, the energy landscape can be much more complex and the barrier considerably reduced. Nevertheless, the Néel-Brown relaxation can be applied also in these cases [15,16,17].

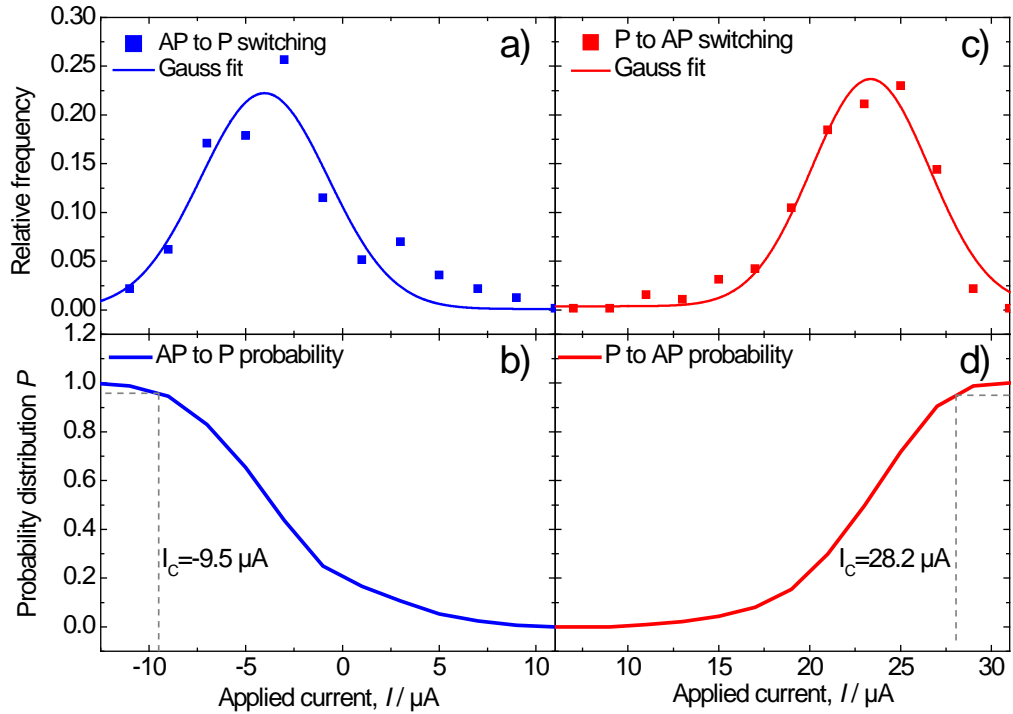

**Figure 5:** Switching statistics of a PMA MTJ with a diameter of 360 nm ($Co_{20}Fe_{60}B_{20}$ (1.0 nm) / MgO (0.84 nm) / $Co_{20}Fe_{60}B_{20}$ (1.2 nm)): Relative number of AP→P switching events in a) with corresponding the probability distribution for switching $P$ shown in c). Relative number of P→AP switching events in b) with the corresponding probability distribution for switching $P$ shown in d). The solid line in a) and c) displays the analysis with a Gaussian function and in b) and b) its integral is shown.

Fitting the above equation to our probability distributions, we obtain an average thermal stability of $\Delta = \frac{\Delta_P + \Delta_{AP}}{2} = 9.2 \pm 1.6$. This value is much lower than a $\Delta > 40$ needed for applications [3,4]. The low thermal stability might be a result from the lowered energy barrier for the case of inhomogeneous magnetization reversal. Thus, in future studies a compromise between low switching currents and thermal stability has to be found, e.g. by layer stack optimization, as suggested in Ref. 13, and size optimization while maintaining the small critical currents.

## V. CONCLUSION

We have studied perpendicular MTJs. The CoFeB layers are perpendicularly magnetized for layer thicknesses below 1.33 nm. With the optimized MTJs, critical switching current densities of as low as $J_c = 9.3$ kA/cm$^2$ are observed. We discuss the average thermal stability of these MTJs, deduced from the switching probability distribution, and estimate an average thermal stability of $\Delta = 9.2 \pm 1.6$. Although we have observed a low $\Delta$, the low switching currents of our MTJs make them good candidates for future studies of thermal STT.

## ACKNOWLEDGMENTS

Funding from the German Research Foundation (DFG) through SPP 1538 "Spin Caloric Transport" is acknowledged. Part of this work was supported by the CRC 1073. A.T. is supported by the Ministry of Innovation, Science and Research (MIWF) of North Rhine-Westphalia with an independent researcher grant. G.R. is supported by the DFG through grant RE1052/22-1.